\documentclass[sigconf,pbalance]{acmart}
\usepackage{graphicx}
\usepackage[skip=0pt]{caption}
\usepackage{subcaption}
\usepackage{xcolor}
\usepackage{xspace}
\usepackage{soul}
\usepackage{fvextra}
\usepackage{xparse}
\NewDocumentEnvironment{minted}{O{} g}{%
  \VerbatimEnvironment
  \begin{Verbatim}[fontsize=\scriptsize,breaklines,breakanywhere,frame=lines,numbers=left,numbersep=4pt]%
}{%
  \end{Verbatim}%
}
\usepackage{enumitem}
\usepackage{cleveref}
\usepackage{multirow}
\usepackage{framed}

\definecolor{exercisebgblue}{rgb}{0,  .69,  .941}
\definecolor{pastelviolet}{rgb}{.81,  .82,  .97}
\newcommand{\hamster}{\textrm{Hamster}\xspace}
\newcommand{\aster}{\textsc{aster}\xspace}
\newcommand{\smalltt}[1]{\texttt{\small #1}}

\newcounter{findingcounter}
\newcounter{implicationcounter}

\definecolor{exercisebgblue}{RGB}{220,235,250}
\definecolor{pastelviolet}{RGB}{230,220,245}
\newenvironment{findingbox}[1]{%
  \refstepcounter{findingcounter}%
  \vspace{-4pt}%
  \begingroup
  \setlength{\parindent}{0pt}%
  \setlength{\parskip}{0pt}%
  \setlength{\FrameSep}{2pt}
  \definecolor{shadecolor}{RGB}{220,235,250}
  \begin{snugshade}%
  \textbf{Finding~\thefindingcounter:}%
}{%
  \end{snugshade}%
  \endgroup
  \vspace{-4pt}%
}

\newenvironment{implicationbox}[1]{%
  \refstepcounter{implicationcounter}%
  \vspace{-4pt}%
  \begingroup
  \setlength{\parindent}{0pt}%
  \setlength{\parskip}{0pt}%
  \setlength{\FrameSep}{2pt}%
  \definecolor{shadecolor}{RGB}{230,220,245}%
  \begin{snugshade}%
  \textbf{Implication~\theimplicationcounter:}%
}{%
  \end{snugshade}%
  \endgroup
  \vspace{-4pt}%
}

\AtBeginDocument{%
  }

\copyrightyear{2026}
\acmYear{2026}
\setcopyright{cc}
\setcctype{by-nc-nd}
\acmConference[ICSE-SEIP '26]{2026 IEEE/ACM 48th International Conference on Software Engineering}{April 12--18, 2026}{Rio de Janeiro, Brazil}
\acmBooktitle{2026 IEEE/ACM 48th International Conference on Software Engineering (ICSE-SEIP '26), April 12--18, 2026, Rio de Janeiro, Brazil}
\acmPrice{}
\acmDOI{10.1145/3786583.3786857}
\acmISBN{979-8-4007-2426-8/2026/04}

\begin{document}

\title{Hamster: A Large-Scale Study and Characterization of Developer-Written Tests}

\author{Rangeet Pan}
\authornote{Both authors contributed equally to this research.}
\email{rangeet.pan@ibm.com}

\affiliation{%
  \institution{IBM Research}
  \city{Yorktown Heights}
  \state{NY}
  \country{USA}
}

\author{Tyler Stennett}
\authornote{Author was an intern at IBM Research at the time of this work.}
\authornotemark[1]
\email{tyler.stennett@gatech.edu}
\affiliation{%
  \institution{Georgia Institute of Technology}
  \city{Atlanta}
  \state{GA}
  \country{USA}
}
\author{Raju Pavuluri}
\email{pavuluri@us.ibm.com}

\affiliation{%
  \institution{IBM Research}
  \city{Yorktown Heights}
  \state{NY}
  \country{USA}
}

\author{Nate Levin}
\email{nlevin33@gatech.edu}
\affiliation{%
  \institution{Georgia Institute of Technology}
  \city{Atlanta}
  \state{GA}
  \country{USA}
}

\author{Alessandro Orso}
\email{orso@uga.edu}
\affiliation{%
  \institution{University of Georgia}
  \city{Athens}
  \state{GA}
  \country{USA}
}

\author{Saurabh Sinha}
\email{sinhas@us.ibm.com}
\affiliation{%
  \institution{IBM Research}
  \city{Yorktown Heights}
  \state{NY}
  \country{USA}
}

\renewcommand{\shortauthors}{Pan et al.}

\begin{abstract}
Automated test generation (ATG), which aims to reduce the cost of manual test suite development, has been investigated for decades and has produced countless techniques based on a variety of approaches: symbolic analysis, search-based, random and adaptive-random, learning-based, and, most recently, large-language-model-based approaches. However, despite this large body of research, there is still a gap in our understanding of the characteristics of developer-written tests and, consequently, our assessment of how well ATG techniques and tools can generate realistic and representative tests. To bridge this gap, we conducted an extensive empirical study of developer-written tests for Java applications, covering 1.7 million test cases from open-source repositories. Our study is the first of its kind to evaluate aspects of developer-written tests that are mostly neglected in the existing literature---including test scope, test fixtures and assertions, types of inputs, and use of mocking---and characterize tests accordingly. Based on this characterization, we then compare existing tests with those generated by two state-of-the-art ATG tools. Our results highlight that the vast majority of developer-written tests exhibit characteristics that are beyond the capabilities of current ATG tools. Finally, based on our findings, we identify promising research directions that can help develop more effective tool support for developer testing practices. We believe this work can set the stage for additional research and bring ATG tools closer to generating the types of tests developers write.
\end{abstract}

\ccsdesc[500]{Software and its engineering~Software testing and debugging}

\keywords{testing, empirical, large language model, developer}



\maketitle

\vspace{-5pt}

\section{Introduction}
\label{sec:intro}

Automated test generation (ATG) is a long-standing and active research topic in the software engineering field. This area has produced an extensive body of work spanning a variety of techniques, including symbolic approaches (e.g.,~\cite{clarke1976testing, king1976symbolic, howden1976symbolic, visser:2004, cadar:2008:exe, cadar:2008:klee, godefroid2005dart, sen2005cute, xie:2005, tillmann2008pex}), search-based methods (e.g.,~\cite{tonella:2008, harman:2010:tse, mcminn:2004, fraser2011evosuite, lin:2021, lukasczyk2022pynguin}), random and adaptive-random techniques (e.g.,~\cite{pacheco2007feedback, ciupa:2008:icse, chen:2010:jss, lin:2009:ase, arcuri:2011:issta, lukasczyk:2023:emse}), fuzzing approaches (e.g.,~\cite{lemieux2018perffuzz, padhye2019jqf, padhye2019zest}), learning-based methods~\cite{durelli2019study}, and, most recently, techniques leveraging large language models (LLMs) (e.g.,~\cite{tufano2020unit, schafer2023empirical, siddiq2024using, tang:2024:tse, ryan2024code, xia2024fuzz4all, gu2025llm, wang2024survey, liu2025llm, vikram2023can, bareiss2022code, pizzorno2024coverup, dakhel2023effective, alshahwan2024automated, pan2025aster}). 

Existing ATG techniques have achieved considerable success in terms of code coverage and mutation scores, with several empirical studies reporting that tool-generated test suites can even surpass developer-written ones in these dimensions~\cite{fraser:2015:developerstudy}. Recent LLM-based approaches further push this boundary by generating tests that are not only effective in terms of coverage, but also more readable, mimicking the stylistic qualities of human-written tests. However, we argue that these metrics and successes capture only a small piece of what makes a test valuable in practice. Coverage and fault detection provide an immediate, quantifiable view of effectiveness, but they do not address the deeper qualities that determine whether tests remain useful and relevant as a system evolves.

In real-world development, tests are not always disposable artifacts generated for a single snapshot of a codebase. Rather, they are often an integral part of the project's quality assurance infrastructure, persisting across multiple iterations of the system and verifying various components. To fulfill this role, tests must have qualities beyond good coverage and fault detection: they need to be understandable, maintainable, and aligned with the structural and behavioral patterns that developers favor. This is especially important in light of the shifting role of ATG.
For years, ATG tools were mostly used to generate regression tests on demand—quick, disposable artifacts that required little long-term attention. With the rise of the new generation of LLM-based coding assistants
(e.g.,~\cite{copilot, watsonx, amazonq, cursor}),
ATG is increasingly being incorporated in a very different way: to help developers generate the core tests they will keep, maintain, and evolve alongside their codebases.

\begin{table}[t]
\caption{Overview of the \hamster dataset.}
\centering
\resizebox{\columnwidth}{!}{
\begin{tabular}{lrrrrrr}
\toprule
& \textbf{Total} & \textbf{Average} & \textbf{P25} & \textbf{P50} & \textbf{P75} & \textbf{P90} \\
\midrule
Projects & 1,908 & -- & -- & -- & -- & -- \\
Application classes & 1,710,445 & 896 & 109 & 368 & 936 & 2274 \\
Application methods & 10,445,194 & 5474 & 514 & 1883 & 5474 & 14577 \\
Test classes & 281,019 & 147 & 8 & 38 & 143 & 388 \\
Test methods & 1,697,196 & 889 & 28 & 186 & 786 & 2297 \\
Test fixture methods & 224,137 & 117 & 1 & 14 & 79 & 296 \\
Test method NCLOC & 37,741,666 & 22 & 6 & 12 & 23 & 42 \\
\bottomrule
\end{tabular}
}
\label{tab:dataset}
\end{table}

This motivates a deeper investigation into the characteristics of developer-written tests and how they differ from those produced by ATG tools. To that end, we take a first step toward bridging this gap by systematically studying the structure and practices of developer-written tests, with a focus on the Java language. Specifically, we mined popular Java projects from GitHub, applying a set of filtering criteria (\S\ref{sec:dataset}), to create a large-scale dataset called Hamster. The dataset consists of 1,908 Java projects, 1,710,445 application classes, 281,019 test classes, and 1,697,196 test methods, providing significant breadth for analyzing the ways developers design tests. Table~\ref{tab:dataset} summarizes key characteristics of the dataset.

We analyzed the test cases in our dataset to understand their characteristics along dimensions both underexplored in prior work and relevant in our own experience. Specifically, we examined five aspects: (1) test scope, (2) test inputs, (3) test fixtures, (4) invocation sequences, and (5) test coverage. 
Test scope was defined as the focal methods exercised by a test, identified through heuristics on class instantiation and invocation. Test inputs were classified based on whether they came from file-system resources or structured data formats (e.g., XML, JSON, CSV). For test fixtures, we measured size and cyclomatic complexity, and analyzed mocking practices by type and number of mocked resources. Invocation sequences—ordered chains of method calls, constructor calls, and assertions—were segmented into behavior blocks followed by assertions, allowing us to measure sequence length, assertion density, and advanced assertion patterns. Finally, coverage was evaluated by instrumenting a subset of projects and collecting coverage data from executed tests.
We assessed two prominent ATG tools: EvoSuite~\cite{fraser2011evosuite, evosuite}, due to its widespread use and recognized effectiveness, and
\aster~\cite{pan2025aster}, as a recent LLM-based approach. 

Overall, our results show that a majority of developer-written tests exhibit characteristics—often across multiple dimensions—that remain beyond the capabilities of current ATG tools. These observations provide insights into the fundamental differences between developer-written tests and those generated automatically. For instance, developer-written tests frequently exercise multiple classes and methods, while ATG-generated tests typically target only a single method, limiting their applicability in scenarios that require more complex testing logic. Developers also tend to rely on complex inputs read from external sources such as files. These inputs often have a rich semantic structure and encode domain-specific information, which existing ATG tools fail to capture. While ATG techniques can generate structured inputs such as dictionaries, they generally lack the ability to encode the semantic meaning that developers incorporate into their test data. In addition, developer-written tests often make extensive use of fixtures, which can be fairly complex and frequently involve mocking both library and application classes. By contrast, ATG-generated tests tend to have either very simple fixtures or none at all. Our paper also discusses the implications of these findings for future research in ATG and outlines possible directions for advancing ATG techniques so that they can generate more realistic tests that not only resemble those written by developers but also align with the needs of modern coding assistants.

The key contributions of this work include:

\begin{itemize}[leftmargin=*]

\item The first-of-its-kind systematic study of the characteristics of developer-written test cases on a large corpus of real-world Java applications.

\item A comprehensive analysis and discussion of the characteristics of developer-written tests along different dimensions, the performance of representative ATG tools in this context, and the implications for future research.

\item An artifact consisting of a large dataset (1,908 real-world Java applications with 1.7M test cases consisting of 37.7M NCLOC) and analysis tools that enable other researchers to replicate and build upon our work~\cite{hamster:artifact}.

\end{itemize}

We see this work as an important first step toward a better understanding of both developer-written and ATG-generated tests, providing evidence that motivates further research in this direction. In particular, our findings suggest that the next generation of ATG tools must go beyond maximizing coverage or fault detection. They should also aim to generate tests that developers can integrate, maintain, and evolve as part of their long-living test suites. 
Our work does not assume that developer-written tests are inherently superior to tool-generated ones in every aspect; rather, it reflects the practical reality that tests aligned with existing test suites and developer practices are more likely to be adopted and maintained.
\section{The Hamster Dataset}
\label{sec:dataset}

The creation of the \hamster dataset started with mining GitHub for popular Java projects (\S\ref{subsec:datasetcollection}). We then analyzed these projects using the JavaParser~\cite{javaparser} and Tree-sitter~\cite{treesitter} through CLDK~\cite{krishna:2025:cldk, cldk} APIs. 
Finally, we used the analysis information to construct the \hamster model, illustrated in Figure~\ref{fig:hamstermodel}.

\subsection{Dataset Collection}
\label{subsec:datasetcollection}
\begin{figure}
    \centering
    \includegraphics[width=\linewidth]{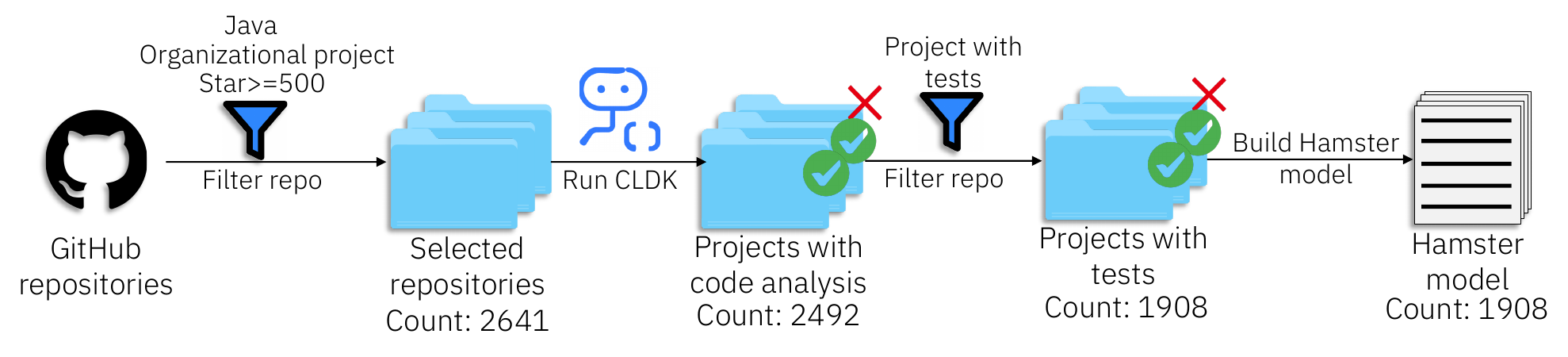}
    \caption{The Hamster data collection process.} 
    \label{fig:datacollection}
\end{figure}

Figure~\ref{fig:datacollection} illustrates the dataset collection process. We began by selecting GitHub repositories using various filtering criteria. We focused on projects written in Java, as that was the scope of our study. To prioritize repositories with high code quality, we included repositories maintained by organizations, excluding those authored by individual developers. We also applied a popularity threshold~\cite{borges2016understanding}, selecting repositories with at least 500 stars to ensure broad usage and relevance. After applying these filters, we obtained 2,641 repositories. We then analyzed these repositories with JavaParser~\cite{javaparser} and Tree-sitter~\cite{treesitter}; the analysis failed for 149 projects (due to unavailable or custom dependencies), which yielded 2,492 projects with analysis information. Finally, we excluded projects that did not have any tests, resulting in the final dataset of 1,908 projects, for which we constructed the \hamster model.

\subsection{Hamster Model Construction}
\label{subsec:hamstermodel}

\begin{figure}
    \centering
    \includegraphics[width=0.95\linewidth]{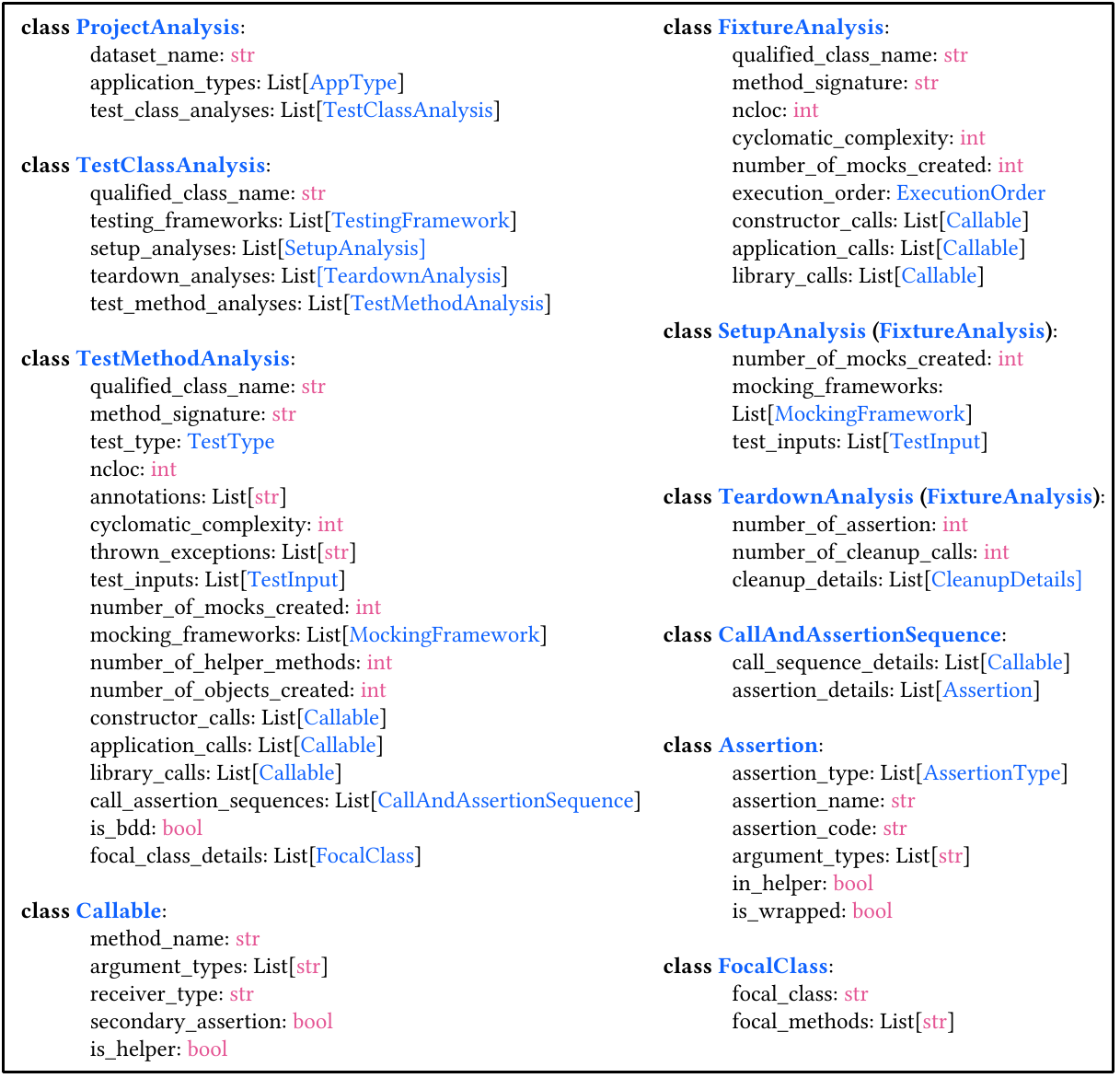}
    \caption{The Hamster analysis data model.} 
    \label{fig:hamstermodel}
\end{figure}

The \hamster model (Figure~\ref{fig:hamstermodel}) represents a rich set of information about Java test cases, characterizing test cases quantitatively in different ways and enabling various test dimensions to be studied. The model can be viewed as consisting of three primary levels---the project, class, and method level---each consisting of their own relevant details. The model is rooted at the \texttt{\small ProjectAnalysis} type, which stores all the information required at the project level, including the project name, its application-type categories, and analysis objects for the test classes of the project. For each test class, through \texttt{\small TestClassAnalysis}, the model stores information about test fixtures (i.e., setup and teardown methods) and test methods. Lastly, \texttt{\small TestMethodAnalysis} models information about test methods and includes details such as lines of code, mocking details, information about call sites, assertions, focal classes and methods, etc. 

\vskip 2pt
\noindent\textit{Project Analysis.}
We tag each project into one or more application types based on the imports in the application compilation units. For this, we used a broad categorization of Java applications into five types---web application, web API, Android, Java EE, and Java SE---and listed relevant library classes for these types. We identified test classes in a project by checking whether a class contains any test methods, based on conventions used in various Java testing frameworks for writing test methods. Thus, we checked for annotations, such as 
\texttt{\small @Test}, \texttt{\small @ParameterizedTest}, \texttt{\small @TestFactory}, \texttt{\small @RepeatedTest}, etc., that indicate test methods. We also identified tests using naming/signature conventions; e.g., in JUnit 3, any public instance method that has no parameters and whose name has the prefix \texttt{\small test} is considered a test method. 

\noindent\textit{Test Class Analysis.}
This analysis extracts information from each test class, including its name, testing frameworks, fixtures, and test methods. Frameworks are identified by analyzing imported types. 
Next, we discuss the key analysis components, which includes analysis of test methods, test fixtures, and call/assertion sequences.

\vskip 2pt
\noindent\textit{Test Method Analysis.}
For each test method, we collect metrics such as NCLOC, cyclomatic complexity, the number and types of mocks created, the mocking frameworks used, the number of objects created, details about constructor, application, and library calls, and assertions made. 
The analysis also handles calls to helper methods, which we define as methods within the same test class that are called within the test method. To capture their contribution, we construct a call graph and integrate their analysis results back into the main test method, ensuring a comprehensive representation of the test's behavior.
We also identify test scope, which refers to the focal classes and methods that the test intended to exercise and validate. To perform our test-scope analysis, we constructed a custom algorithm based on the creation and usage of objects within a test as discussed in \S\ref{subsec:testscope}. 

\vskip 2pt
\noindent\textit{Test Fixture Analysis.}
Our test fixture analysis focuses on setup and teardown methods that, respectively, prepare the environment before a test runs and clean up resources afterward. We detect test fixtures by combining multiple signals: method naming conventions (e.g., \smalltt{setUp()} and \smalltt{tearDown()} in JUnit 3), method-level annotations (e.g., \smalltt{@Before} and \smalltt{@After} in JUnit 4 and 5), inheritance patterns, and class-level annotations. To ensure inherited behavior is captured, we construct an inheritance graph of the test class and identify any public or protected fixtures inherited from parent classes.

For characterizing setup methods, we extract general method information, such as NCLOC and cyclomatic complexity, with particular attention to the creation and configuration of mocked resources. Teardown methods, in contrast, are analyzed for both resource cleanup behavior and embedded assertions that confirm successful completion. Our analysis highlights resource types that are typically managed during teardown, such as I/O streams, network connections, and database handles. To recognize cleanup operations, we apply heuristics based on method names (e.g., \smalltt{close()}, \smalltt{terminate()}) and match receiver objects against a curated list of known libraries to infer the type of resource being released. 

\vskip 2pt
\noindent\textit{Call and Assertion Sequence Analysis.}
For each test method, we construct an invocation sequence, defined as the complete ordered list of callable entities---method and constructor calls, method references (e.g., \smalltt{String::isEmpty}), and assertions---preserved in their runtime order. This ordering ensures that argument expressions are resolved before their enclosing call, lambda expressions are expanded, anonymous subclass instantiations with initializer blocks are processed, and chained method invocations are interpreted from left to right.

Each callable entity is classified as either an assertion or a non-assertion, based on mappings from widely used testing frameworks and assertion libraries. Assertions are further categorized into types such as truthiness, equality, identity, nullness, and numeric tolerance, following a taxonomy derived from framework documentation and summarized in our artifact~\cite{hamster:artifact}.

To analyze test structure, we partition each invocation sequence into call-assertion sequences. A call-assertion sequence begins with a contiguous run of non-assertion entities (the call sequence), followed by the assertions that immediately succeed them (the assertion sequence). A new call-assertion sequence then starts at the next non-assertion. This partitioning exposes how developers interleave state-manipulating actions with verification steps, and allows for further analysis.

\subsection{Dataset Characteristics}
\label{subsec:characteristics}

Overall, we constructed the \hamster model for 281K test classes, 1.7M test methods, and 224K test fixture methods, analyzing over 37.7M lines of test code in the process.

\begin{figure}[t]
    \centering
        \includegraphics[width=\columnwidth]{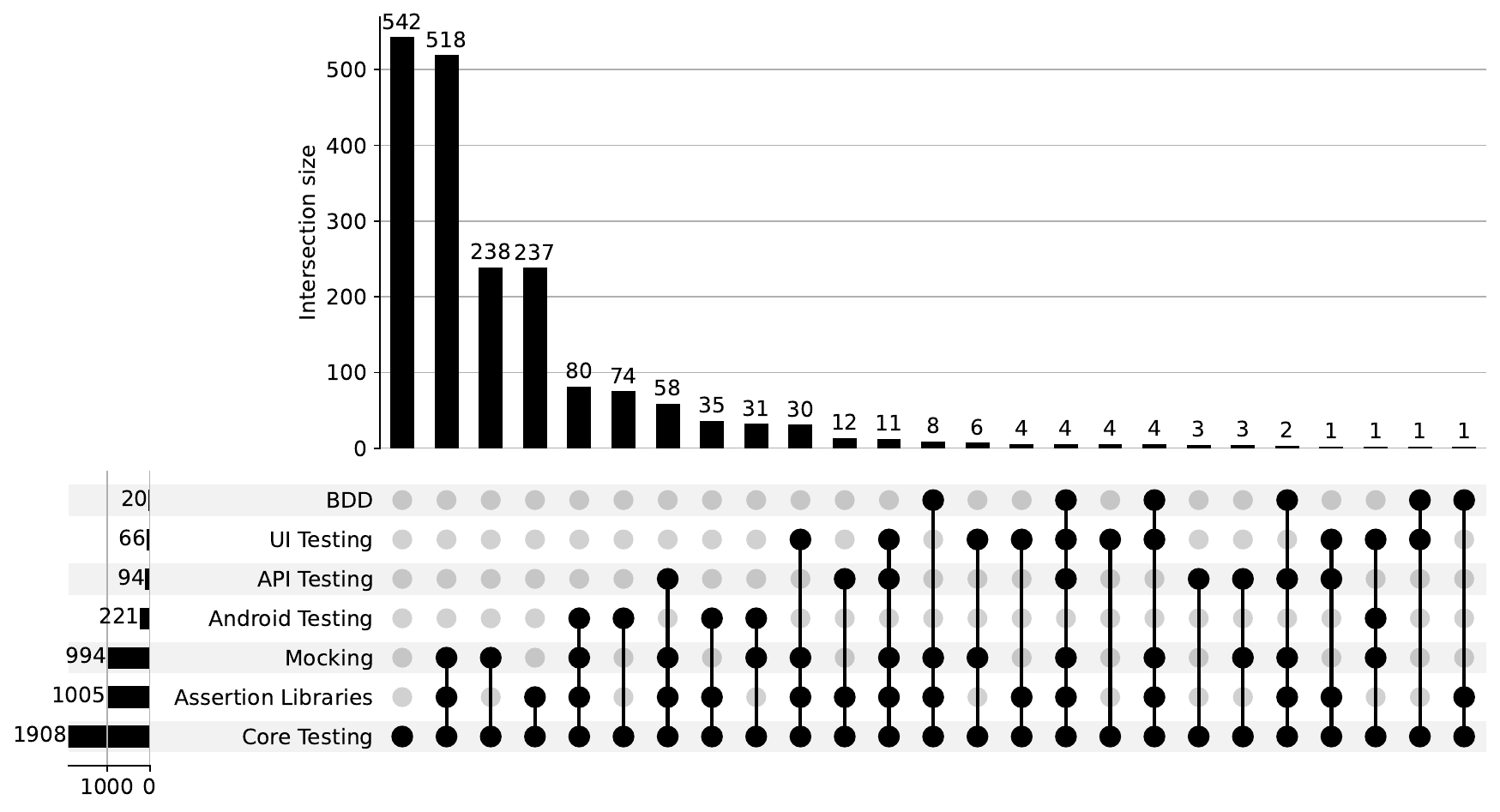}
    \caption{Usage of different categories of testing frameworks in the dataset.}
    \vspace{-5pt}
    \label{fig:testingframework}
\end{figure}

Figure~\ref{fig:testingframework} illustrates the distribution of different types of testing frameworks used by the applications in the \hamster dataset. We group the testing frameworks into seven categories, which in decreasing order of their usage counts across projects are: core frameworks (1,908), assertion libraries (1,005), mocking frameworks (994), Android testing frameworks (221), API testing frameworks (94), UI testing frameworks (66), and BDD frameworks (20). About 28\% ($542/1908$) of the applications use only core testing frameworks, which consist of JUnit and TestNG. Thus, a vast majority of the applications use at least two types of testing frameworks. Among these, 611 applications use three different types of frameworks, while another 178 use four types of frameworks; finally, 21~applications use five or more types of testing frameworks.

\section{Research Questions}
\label{sec:rq}


We formulated a set of research questions (RQs) to understand various characteristics of developer-written tests. 

\vspace{-4pt}
\begin{itemize}[leftmargin=*]
    
    \item \textbf{RQ1 (Test fixture):} Are test fixtures used commonly in developer-written tests and what are the unique patterns in test fixtures?
    \item \textbf{RQ2 (Test input):} What are the different categories of inputs used in tests and how complex are they?
    \item \textbf{RQ3 (Test scope):} What are the different test scopes and how prevalent are they in developer-written tests?
    \item \textbf{RQ4 (Test method characteristics):} What are the characteristics of developer-written test methods, considering both their structural complexity (e.g., NCLOC, object constructions, method calls) and the organization of call-assertion sequences?

    \item \textbf{RQ5 (Code coverage):} How effective are developer-written tests in terms of code coverage?
\end{itemize}

\subsection{RQ1: Characteristics of Test Fixtures}
\label{subsec:testfixture}
\noindent
\textit{Characteristics of test fixtures.}
Analyzing fixtures across all test classes, our dataset contains 172,099 setup methods and 52,038 teardown methods. As seen in Table~\ref{tb:fixture}, we assess fixture complexity using NCLOC, cyclomatic complexity, and operational breadth, which we approximate through the number of objects instantiated, mocks created, and cleanup operations performed.

Setup methods initialize shared resources and objects. The following fragment shows a simple setup that initializes base URLs for subsequent tests.

\vspace{-3pt}
\begin{minted}[frame=lines,framesep=1mm,baselinestretch=1, fontsize=\scriptsize, breaklines, breakanywhere, linenos,numbersep=4pt]{java}
public void setUp() throws Exception {
    this.propertiesUrl1 = new URL("http://localhost:" + port + "/properties/1");
    this.propertiesUrl2 = new URL("http://localhost:" + port + "/properties/2");}
\end{minted}
\vspace{-3pt}

Teardown methods can vary widely in complexity, from simple object cleanup to sophisticated operations such as database rollbacks. For example, in Apache ActiveMQ Artemis~\cite{artemis}, test classes inherit from a common base that launches an MQ server and configures per-test state. The corresponding teardown not only releases resources but also asserts correct cleanup. Combined with the inherited logic, this teardown exceeds 100 lines of code with cyclomatic complexity of 16.

\vspace{-3pt}
\begin{minted}[frame=lines,framesep=1mm,baselinestretch=1, fontsize=\scriptsize, breaklines, breakanywhere, linenos,numbersep=4pt]{java}
public class DuplicateRecordIdTest extends ActiveMQTestBase {
   public void setUp() throws Exception {
      super.setUp();
      ...}...
}      
public abstract class ActiveMQTestBase extends ArtemisTestCase{ ...
   @BeforeEach
   public void setUp() throws Exception {
      sendMsgCount = 0;...}
   @AfterEach
   public void tearDown() throws Exception {
      try {
         closeAllSessionFactories();
         ...
         assertAllClientSessionsAreClosed();
      } finally {
            + 95 lines ...} } }
\end{minted}
\vspace{-3pt}

\begin{table}[t]
\caption{Test fixture characteristics.}
\centering
\resizebox{\linewidth}{!}{
\begin{tabular}{l|c|c|c|c|c|c|c|c}
\toprule
& \multicolumn{4}{c|}{\textbf{Setup}} & \multicolumn{4}{c}{\textbf{Teardown}}\\
\cmidrule{2-9}
\textbf{Metrics} & \textbf{P25} & \textbf{P50} & \textbf{P75} & \textbf{P90}& \textbf{P25}& \textbf{P50} & \textbf{P75} & \textbf{P90}\\
\midrule
NCLOC & 4 & 7 & 16 & 37& 3 & 4 & 6 & 12\\
Cyclomatic complexity & 1 & 1& 2&5& 1 & 1 & 2 & 3\\
\# of objects created & 0 & 1 & 2 & 5& - & - & - & -\\
\# of cleanup operations & - & - & - & -& 0 & 0 & 1 & 1\\
\# of mocks created & 1& 2& 3& 7&-& -& -&- \\
\bottomrule
\end{tabular}
}
\label{tb:fixture}
\end{table}
\noindent
\textit{Modularity and reuse.} Test fixtures promote modularity by encapsulating setup and teardown logic, clearly separating environment preparation and cleanup from the actual test behavior. This separation improves readability, maintainability, and scalability. Fixtures also enable reuse: common setup routines can be centralized across test classes, reducing duplication. In practice, this reuse is often coupled with inheritance, where test class hierarchies share initialization logic while allowing module-specific customization.


The Netflix Maestro~\cite{maestro} project illustrates this pattern, where \smalltt{MaestroBaseTest} defines a shared skeleton, \smalltt{MaestroEngineBaseTest} adds module-specific initialization, and concrete classes such as \smalltt{StepRuntimeSummaryTest} (shown below) inherit shared fixture logic while adding their own state.

\vspace{-3pt}
\begin{minted}[frame=lines,framesep=1mm,baselinestretch=1, fontsize=\scriptsize, breaklines, breakanywhere, linenos,numbersep=4pt]{java}
/** Test setup **/
public class StepRuntimeSummaryTest extends MaestroEngineBaseTest
  @BeforeClass
  public static void init() {
    MaestroEngineBaseTest.init();}
  . . .  }
/** Common fixture methods across several test methods **/
public abstract class MaestroEngineBaseTest extends MaestroBaseTest {
  /** start up. */
  @BeforeClass
  public static void init() {
    MaestroBaseTest.init();
    ... }
  @AfterClass
  public static void destroy() {
    evaluator.preDestroy();
    MaestroBaseTest.destroy();}}
/** Test base class. */
public class MaestroBaseTest {
  @BeforeClass
  public static void init() {}
  @AfterClass
  public static void destroy() {}...}
\end{minted}
\vspace{-2pt}

These examples illustrate how fixtures not only enhance modularity by isolating test concerns, but also provide a foundation for reuse across complex test suites. Simultaneously, the reliance on layered inheritance and shared states can introduce substantial complexity during test design.

\vspace{3pt}
\begin{findingbox}{3}
{
Test fixtures are widely used in developer-written tests, with 43.1\% and 16.2\% of test classes containing at least one setup and teardown method, either defined or inherited, respectively. Fixtures play a critical role in enabling modular test suite design, supporting isolation of functionality while reusing shared components.
}
\end{findingbox}
\vspace{3pt}

\noindent
\textit{Mocking.} Test fixtures are often used to manage mocks for various resources. In our dataset, 9.8\% of the tests contain some mocking code in the test setup method.
The following example from Apple ServiceTalk~\cite{appleservicetalk}, for instance, illustrates a case where the setup method is used to create the HTTP resources needed to run the tests in a mocked environment.
\begin{minted}
[frame=lines,framesep=1mm,baselinestretch=1, fontsize=\scriptsize, breaklines, breakanywhere, linenos,numbersep=4pt]{java}
public void setUp() {
    mockRequestMetadata = mock(HttpRequestMetaData.class);
    when(mockRequestMetadata.headers()).thenReturn(mock(HttpHeaders.class));
        + 9 lines ...}
\end{minted}

\vspace{2pt}
\begin{findingbox}{4}
{Test fixtures frequently use mocking to create an isolated execution environment.}
\end{findingbox}
\vspace{3pt}

\noindent
\textit{Execution pattern.}
Most testing frameworks let developers control the execution order of setup and teardown methods, typically through annotations or configuration directives. In our dataset, 64.9\% of the fixture methods are configured to run before or after every test method, with 27.3\% executed only once at the class level, either before or after all test methods in the class are run. Tests using both types of fixture executions also occur in practice, as illustrated by the following example from the Epoxy project~\cite{eproxy}. 
The \smalltt{beforeClass} method initializes shared fields that are used in multiple test methods. Before each test, the \smalltt{setUp} method is executed to register a \smalltt{TestObserver} object, allowing the tests to probe UI elements independently and ensure test isolation by avoiding shared mutable state. Finally, after all test methods in the class have executed, the \smalltt{afterClass} method runs to report summary statistics about the test actions.

\begin{minted}
[frame=lines,framesep=1mm,baselinestretch=1, fontsize=\scriptsize, breaklines, breakanywhere, linenos,numbersep=4pt]{java}
public class DifferCorrectnessTest {...
  @BeforeClass
  public static void beforeClass() {
    totalDiffMillis = 0;
    totalDiffOperations = 0;
    totalDiffs = 0;}
  @AfterClass
  public static void afterClass() {...
    System.out.println("Average operations per diff: " + avgOperations);}
  @Before
  public void setUp() {...
  testAdapter.registerAdapterDataObserver(testObserver);}}
\end{minted}
\vspace{-2pt}

\noindent
\textit{Cleanup patterns.} While setup methods are essential for preparing the test environment, teardown methods play a critical role in making tests more self-contained and reliable. The primary responsibility of a teardown method is to clean up any resources initialized during the setup phase, ensuring that tests do not leave residual state that could interfere with other tests, and resources are properly released to prevent issues such as resource leaks, resource contention, or security vulnerabilities.
Teardown behavior is highly dependent on the types of resources used in the test (e.g., server, socket, database, file system, or other external resources) and can be complex, as illustrated earlier by the ActiveMQ Artemis example. The following examples show common patterns of teardown methods for various resource types.

\vspace{-2pt}
\begin{minted}
[frame=lines,framesep=1mm,baselinestretch=1, fontsize=\scriptsize, breaklines, breakanywhere, linenos,numbersep=4pt]{java}
// Stopping a server 
  public void stopServer() {
    server.stop().join();
    ...}
// Closing a socket
  public void tearDown() throws Exception {
    if (mSocket != null) mSocket.close();
    if (mServerSocket != null) mServerSocket.close();
    super.tearDown();}
// Closing database connection
  public static void stop() {...
    if (dataSource != null) {
      try {
        dataSource.close();
      } catch (SQLException e) { }}
// Closing files
  public void tearDown() {...
    fs.close();}
\end{minted}
\vspace{-2pt}

\noindent
\textit{Comparison with ATG Techniques.}
Tests generated by tools such as EvoSuite and \aster rarely contain fixtures. 
Tests generated by these tools also lack modularity and reuse. They neither enforce structured execution environments nor leverage reusable patterns within the application under test, resulting in limited maintainability compared to manually written fixtures. While their primary focus remains maximizing coverage and fault detection, they often miss the code structures that support long-term test evolution.

\vspace{3pt}
\begin{findingbox}{5}
{Developers often write sophisticated fixtures incorporating various patterns, whereas ATG-generated tests rarely have fixtures or contain very simple fixtures without the typical developer patterns, making the tests less maintainable.}
\end{findingbox}
\vspace{3pt}

\noindent

\textit{Implications.}
One of our key findings is how developers incorporate modularity into their test suites. Although modularity has been extensively studied in the context of software design, surprisingly little work has addressed modularity in test generation. Some prior work has explored modularity in UI testing (e.g.,~\cite{yandrapally2015automated, stocco:2017:apogen}), but a significant gap remains for modular test generation. From the standpoint of ATG techniques, creating test suites with a modular structure presents a promising research direction. Future tools could aim to either refactor existing test suites to make them modular or generate modular tests by identifying and extracting reusable components, such as shared fixture methods.

While we acknowledge the general fixture test smell---where overly broad or unfocused fixtures can reduce clarity and maintainability~\cite{van2001refactoring}---this does not diminish the importance of fixtures overall. Instead, it highlights the need for balance: ATG methods should support the use of fixtures as a means of structuring and reusing test logic, but without encouraging practices that lead to unnecessarily complex or over-generalized setups.

Better mocking support in tool-generated tests that reflect developer concerns with creating isolated test environments is another important research direction. Such capability could be grounded in user-provided test intent. Together, such advances can make tool-generated test fixtures more realistic and sophisticated, bringing them closer to the types of fixtures developers write, while remaining mindful to avoid common fixture-related smells.

\vspace{2pt}
\begin{implicationbox}{2}
{
An interesting direction for future research is the generation of tests in a more modular fashion by creating reusable components---starting from using more fixture methods---so that ATG-generated tests are more readable, easier to maintain, and in-line with developer expectations. Another direction is enhanced support for mocking in tool-generated fixtures.
}
\end{implicationbox}

\subsection{RQ2: Characteristics of Test Inputs}
\label{subsec:testinput}
Test inputs play a key role in test structure. Although test inputs are often defined directly in the test code, many tests rely on structured inputs from external resources. In this research question, we focus on the latter---analyzing the use of structured inputs in tests.

\vskip 1pt
\noindent
\textit{Characteristics of test inputs.}
Our analysis revealed that 72,232 tests (approximately 4.2\%) incorporate at least one structured input within their test scope---including setup methods, test methods, and helper methods. 
The most common input formats confidently identified were classpath-based resources (43.2\%), SQL (29.7\%), and JSON (17.8\%).
A manual review of these instances uncovered several recurring patterns that we discuss in the next paragraphs.

\vskip 1pt
\noindent
\textit{Frequent usage patterns.}
One prominent pattern we observed involves simulating web interactions, where HTML inputs are parsed to evaluate browser outputs or user interface behavior. Similarly, structured formats like JSON and XML are commonly used to mimic HTTP request and response payloads. This approach allows developers to test parsing, validation, and core business logic without relying on actual network communication. For example, in the following test from the Azure Spring Cloud project~\cite{azurespring}, JSON data is loaded from a local file to emulate an API response. This data is then stubbed using Mockito to simulate the body of an incoming HTTP request.
This pattern effectively decouples tests from external services while testing on realistic inputs.

\vspace{-3pt}
\begin{minted}[frame=lines,framesep=1mm,baselinestretch=1, fontsize=\scriptsize, breaklines, breakanywhere, linenos,numbersep=4pt,]{java}
public void validationParsing() {...
    requestBody = mapper.readValue(new File(GET_TEST_REFRESH), JsonNode.class).toString();...
    assertEquals("https://testconfig.azconfig.io", endpoint.getEndpoint());}
\end{minted}
\vspace{-3pt}

Other common patterns include configuration bootstrapping, where property or JSON files externalize parameters such as environment settings, enabling modular and reusable test setups. For instance, tests may load property files to configure mock servers or dependency injection containers.
Structured inputs are also used for database initialization and validation. When tests involve external databases, structured data (e.g., CSV, JSON) helps populate initial state and verify correctness, ensuring tests remain self-contained and reproducible. A representative example is found in the Apache Phoenix project~\cite{apachephoenix}, where a parser processes CSV records and compares them to query results.
\vspace{-4pt}
\begin{minted}[frame=lines,framesep=1mm,baselinestretch=1, fontsize=\scriptsize, breaklines, breakanywhere, linenos,numbersep=4pt,]{java}
public void testTDVCommonsUpsert() throws Exception {...
    PreparedStatement statement = conn.prepareStatement(...);
    ResultSet phoenixResultSet = statement.executeQuery();
    parser = new CSVParser(...);
    for (CSVRecord record : parser) {
        ...
        for (String value : record) {
            assertEquals(value, phoenixResultSet.getString(i + 1));...}}
\end{minted}

\vspace{-3pt}
\noindent
\textit{Structured inputs and test complexity.}
Although structured inputs enhance maintainability and modularity, they can also introduce complexity, as reading and processing external data often requires more intricate code. To understand this tradeoff, we analyzed test complexity based on the presence or absence of structured inputs, considering NCLOC, cyclomatic complexity, and constructor counts. Our analysis revealed that tests with structured inputs exhibit substantially higher complexity: a median cyclomatic complexity of 6 versus 2, and a median length of 51 lines versus 17. This increase likely stems in part from additional object construction---the median number of constructor calls rises from 2 to 5 when structured inputs are present.

\vspace{3pt}
\begin{findingbox}{9}
{
Creating structured inputs in external test resource files helps in writing modular and efficient tests for complex scenarios that involve external resources, databases, etc., but it comes with additional code complexity. 
}
\end{findingbox}
\vspace{3pt}

\noindent
\textit{Comparison with ATG Techniques.} 
The tests generated by ATG tools typically rely on primitive values or simple object hierarchies, with limited use of complex, structured inputs. Although some of these tools are capable of handling advanced scenarios involving databases or external resources, they often fall short in incorporating structured inputs such as JSON, XML, or property files.

\vskip 2pt
\noindent
\textit{Implications.}
Although there is work on exploring the use of structured inputs in ATG tools---for example, G-EvoSuite~\cite{olsthoorn2020generating} extends EvoSuite with grammar-based fuzzing to construct and mutate well-formed JSON inputs, and BRMiner~\cite{ouedraogo2025enriching} employs LLMs to extract realistic inputs from issue reports to enhance EvoSuite~\cite{fraser2011evosuite} and Randoop~\cite{pacheco2007randoop}---a gap remains in understanding how to infer the semantics of tests in order to generate domain-aware structured inputs that are both more relevant to the intended behavior and more reusable. In particular, advancing techniques that can automatically uncover behavior-specific schemas and synthesize corresponding semantic inputs represents a promising research direction.

\vspace{2pt}
\begin{implicationbox}{4}
{
An important research direction is the development of improved ATG approaches that natively support semantic, structured input generation, potentially using LLMs to infer domain-aware schemas from application code.
}
\end{implicationbox}
\vspace{-5pt}
\subsection{RQ3: Identifying Test Scope}
\label{subsec:testscope}

To determine the scope of tests, we first reviewed prior techniques and considered their applicability. Name-based heuristics, such as those proposed by Tufano et al.~\cite{tufano2020unit}, can be effective in simple settings but struggle when multiple focal methods appear in the same test. Other work~\cite{ghafari2015automatically} has used static analysis and data-flow reasoning to trace assertions back to mutator methods. While more systematic, these approaches may miss cases where multiple behaviors are combined into a single oracle, and their reliance on source code for mutator inspection makes them less suited for external library classes---an area that, as we show, is often a subject for tests.

\begin{figure}
    \centering
    \includegraphics[width=0.95\linewidth]{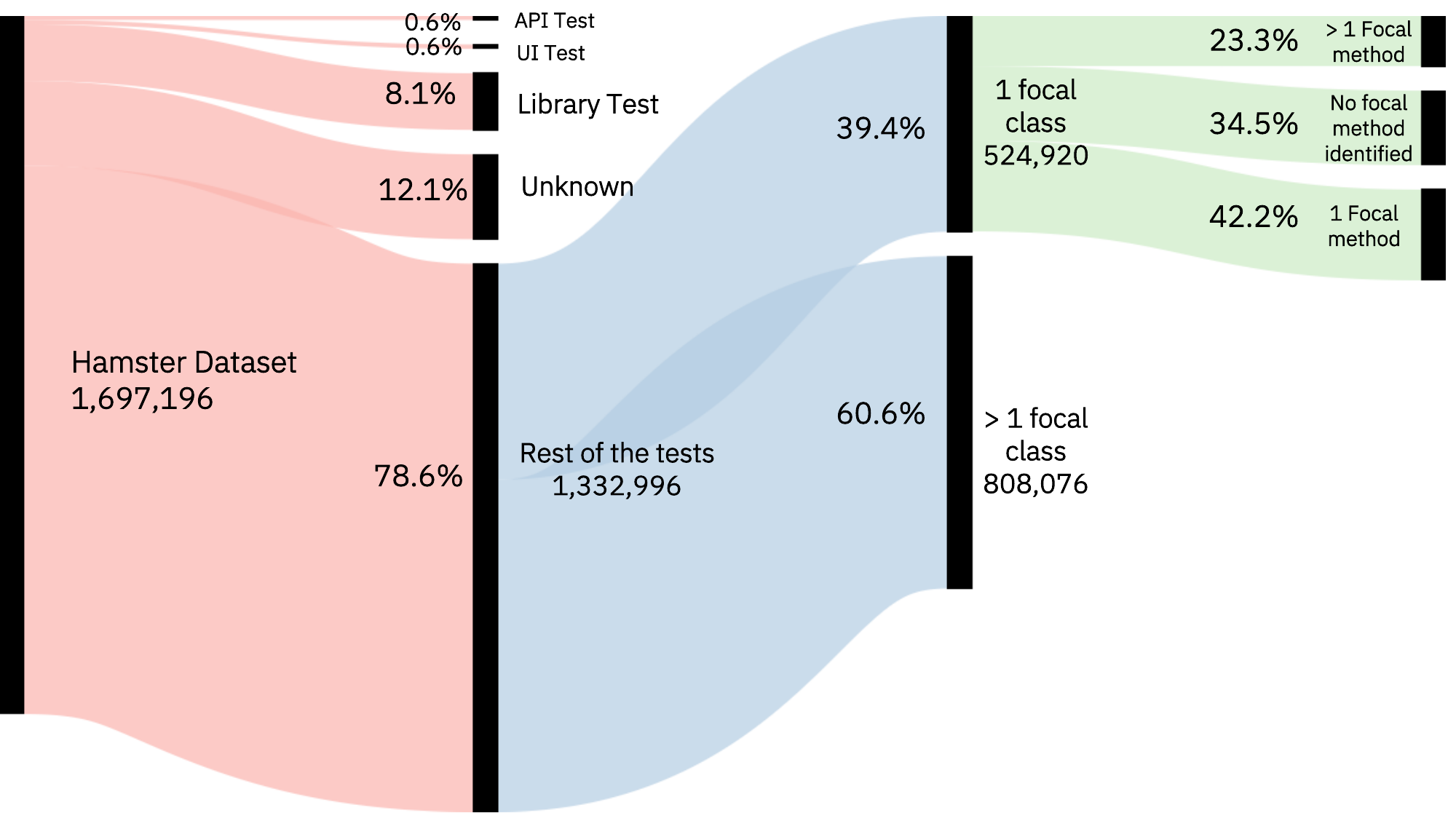}
    \caption{Test-scope analysis for the test methods in the Hamster dataset.} 
    \label{fig:testinglevel}
    \vspace{-7pt}
\end{figure}


To address these limitations, we designed an analysis that infers focal elements directly from object creation and usage within a test while building upon existing approaches. The process examines all objects created in the test body, setup methods, and helper methods, matching constructor invocations to variables to infer concrete types. As the analysis processes each method call, invocations on application classes are tracked under the inferred receiver type when available, or the static type otherwise, excluding getters. Static method calls track the declaring class as focal while linking returned values to variables for factory patterns. Objects serving only as arguments to other application methods are filtered out unless they appear in assertions, and mocked fields are excluded as test doubles. The remaining application classes and their tracked methods constitute the focal scope. Beyond focal identification, our analysis categorizes tests by type: tests referencing UI or API framework types are labeled accordingly, tests with no application focal classes are classified as \textit{library} tests, and tests exercising focal methods are categorized as \textit{unit} or \textit{integration} based on focal class count. Tests with no identifiable focal scope are labeled separately.

\vspace{3pt}
\noindent
\textit{Test scope analysis.} Figure~\ref{fig:testinglevel} illustrates the results of test scope analysis on the 1.7M test methods in the Hamster dataset. UI and API tests account for small fractions of the total tests. After handling those tests, we computed test scopes for the remaining methods. 
Interestingly, 8.1\% of the tests contain no application focal methods and instead target library methods directly. Developers write these tests to verify how library methods behave on predictable inputs in isolation, ensuring that the libraries function correctly when integrated into their applications.
For 12.1\% of the methods, our analysis detected no application or library focal class. On manual examination of a few instances, we found that JavaParser was not able to fully expand the data types to their fully qualified types, on which our analysis is dependent. 
Among the 78.6\% tests that have an application focal class, 39.4\% have one focal class, within which 42.2\% have one focal method, 23.3\% have more than one focal method, and the remaining 34.5\% have no focal methods. For instance, in the example below~\cite{aimacode}, the test is designed to validate that the constructor of the application class \smalltt{Rule} correctly populates its fields. In this case, the focal class is \smalltt{Rule} with no focal method.

\vspace{-3pt}
\begin{minted}
[frame=lines,framesep=1mm,baselinestretch=1, fontsize=\scriptsize, breaklines, breakanywhere, linenos,numbersep=4pt]{java}
public void testANDRule() {
    Rule<Action> r = new Rule<>(new ANDCondition(new EQUALCondition(...));
    Assert.assertEquals(ACTION_EMERGENCY_BRAKING, r.getAction());...}
\end{minted}

\begin{findingbox}{1}
{UI and API tests make up only a small fraction of developer-written tests, whereas 8.1\% target library methods. Overall, 30.9\% and 47.6\% of the tests have one and multiple focal classes, respectively. Among single-focal-class tests, most exercise one focal method.
}
\end{findingbox}
\vspace{3pt}

\begin{figure}[t]
  \centering
  \begin{minipage}{0.5\linewidth}
    \centering
    \begin{subfigure}[t]{0.47\linewidth}
      \centering
      \includegraphics[width=\linewidth]{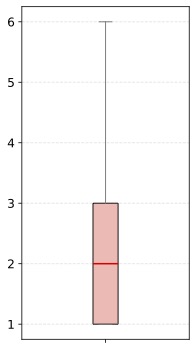}
      \caption{Focal class}
    \end{subfigure}
    \hfill
    \begin{subfigure}[t]{0.49\linewidth}
      \centering
      \includegraphics[width=\linewidth]{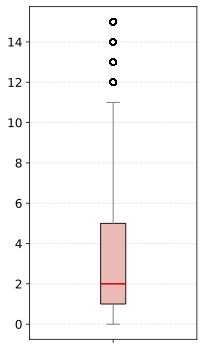}
      \caption{Focal method}
    \end{subfigure}
  \end{minipage}
  \caption{Number of focal classes and methods per test.}
  \label{fig:focal-class-method-distribution}
\end{figure}

\noindent
\textit{Distribution of focal class and method.}
Figure~\ref{fig:focal-class-method-distribution} shows the distribution of focal class and method counts per test (after removing outliers). Our analysis reveals that 39.4\% of the tests in this category target a single focal class, while the remaining 60.6\% involve multiple focal classes. 

\vskip 4pt
\noindent
\textit{Comparison with ATG Techniques.}
EvoSuite~\cite{fraser2011evosuite} and \aster~\cite{pan2025aster} consistently produce tests with exactly one focal class and one focal method as they are designed for unit test generation.
However, when compared to developer-written tests, these tools fall short in scenarios where a test needs to validate multiple focal methods or classes, which is often the case in real-world functional testing.

\vspace{3pt}
\begin{findingbox}{2}
{Developer-written tests often exercise multiple focal classes and methods, whereas tests generated by tools like EvoSuite and \aster consistently target only one focal method. While this makes them effective for isolated unit testing, it limits their applicability in scenarios involving more complex testing logic.}
\end{findingbox}
\vspace{3pt}

\noindent
\textit{Implications.}
An interesting research direction is to create LLMs for test generation. Although there has been some work in this area~\cite{tufano2020unit, rao2023cat, he:2024:unitsyn} focusing on unit testing,
our analysis highlights that developer-written tests often span multiple focal classes and methods, suggesting that future research could (1) explore architectures like mixture-of-experts, where different components of the model specialize in generating tests of different types, and (2) use instructions from developers in natural language to guide the test generation for complex scenarios involving multiple focal classes and methods. There are some recent attempts in the latter direction~\cite{alagarsamy2025enhancing}, but there remains a significant opportunity for improvement---in particular, investigating how tests targeting complex scenarios could be generated automatically from structured specifications or unstructured natural language descriptions, and formally checked for correctness against the specifications. 
\begin{figure*}[t]
    \centering
    \includegraphics[width=0.91\linewidth]{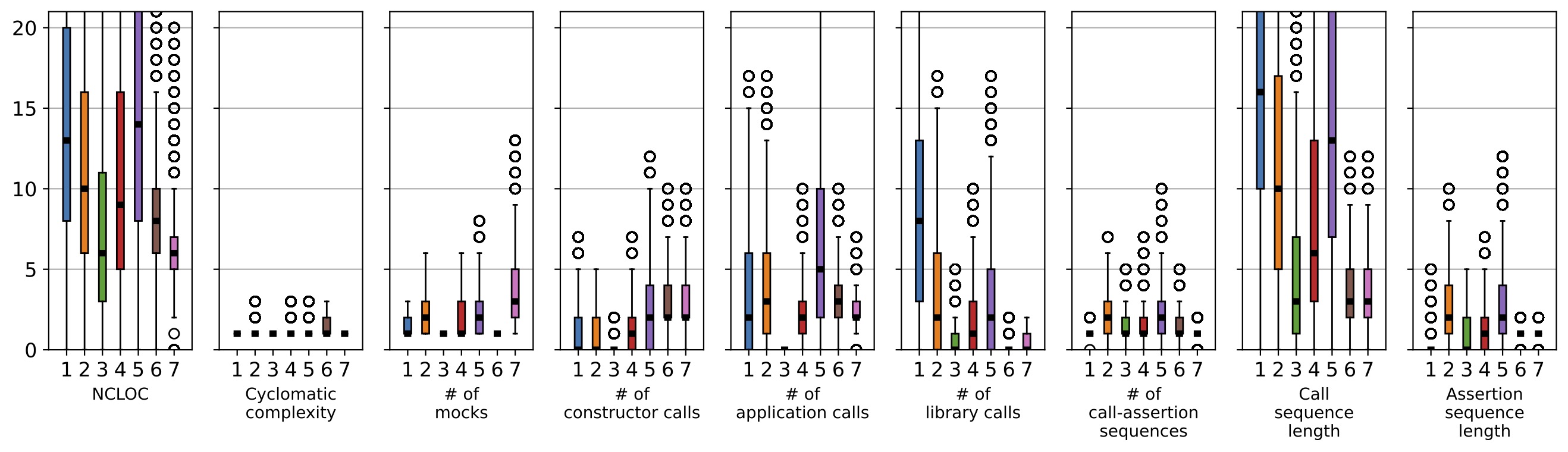}
    \vspace{-2pt}
    \caption{Test method characteristics: 
Developer-written Tests: (1) API Tests, (2) UI Tests, (3) Library Tests, (4) Rest of the tests with one focal class, and (5) Rest of the tests with multiple focal classes. ATG tools: (6) EvoSuite and (7) \aster.\protect\footnotemark}

    \vspace{-8pt}
    \label{fig:testmethod}
\end{figure*}
Another important research direction is to better understand the motivations behind different test scopes in developer-written tests. For instance, do developers tend to favor fine-grained unit tests in the early development stages, where they need to validate each unit in isolation, while their focus shifts toward writing more functional tests as the project matures? Are there developer, project, or organizational preferences in these respects? Investigation of these questions can lead to the development of techniques that effectively assist developers in writing tests at different project stages and enable more tailored support that aligns with both the project context and individual developer preferences.

\vspace{3pt}
\begin{implicationbox}{1}
{Our findings can guide future research in training more specialized test-generation LLMs. Another direction is to understand developer and project preferences and build more tailored test generators for increasing developer productivity.}
\end{implicationbox}

\subsection{RQ4: Characteristics of Test Methods}
\label{subsec:testmethod}

Next, we analyze test methods to extract key characteristics and better understand their structure. We start by examining their complexity and then delve into the sequences of method calls and assertions to identify common testing patterns.

\vskip 2pt
\noindent\textit{Characteristics of test methods.} We analyze test method features across various dimensions, including NCLOC, cyclomatic complexity, the number of objects created, the number of application and library method calls, and call-assertion sequence length. We define a test method as more complex when it exhibits higher values along these dimensions. As shown in Figure~\ref{fig:testmethod}, developer-written tests (the first five categories, separated by test scope) are generally more complex than ATG-generated tests (the last two categories). 
In particular, tests with multiple focal classes consistently exhibit the highest complexity, combining greater NCLOC, numerous object constructions, longer call sequences, longer assertion sequences, and extensive application and library calls. Often developer-written single-class tests exceed the complexity of ATG ones, emphasizing how human-written tests capture richer behaviors and interactions.

\vspace{3pt}
\begin{findingbox}{6}
{
Developer-written tests are substantially more complex than tool-generated ones, involving higher NCLOC, more object creations, greater use of application and library calls, and longer call-assertion sequences. Among these, tests spanning multiple focal classes are consistently the most complex.
}
\end{findingbox}
\vspace{3pt}

\footnotetext{Developer-written tests use the full \hamster dataset. EvoSuite is evaluated on 40 projects, and \aster on tests from the original paper (7 projects, 6 LLMs). Project selection for ATG tools is detailed in \S\ref{subsec:sota}.}

\vskip 2pt
\noindent\textit{Patterns and characteristics of call-assertion sequences.}
We examine developer-written tests using their partitioned invocation sequences, focusing on the resulting call-assertion sequences. 

\noindent
\textbf{Repetitive testing blocks.}
Developers often structure their tests as a sequence of method calls followed by assertions, and frequently repeat these call-assertion blocks with variations in the input parameters. For example, in the NewPipe project~\cite{teamnewpipe}, tests for YouTube services invoke the same extraction method multiple times, each with slightly different parameters, and then assert that the resulting object belongs to the expected type.
\vspace{-5pt}
\begin{minted}[frame=lines,framesep=1mm,baselinestretch=1, fontsize=\scriptsize, breaklines, breakanywhere, linenos,numbersep=4pt, style=xcode]{java}
void getPlaylistExtractorIsMix() throws Exception {
    final String videoId = "_AzeUSL9lZc";
    PlaylistExtractor extractor = YouTube.getPlaylistExtractor(
        "https://www.youtube.com/watch?v=" + videoId + "&list=RD" + videoId);
    assertTrue(extractor instanceof YoutubeMixPlaylistExtractor);
    extractor = YouTube.getPlaylistExtractor(
        "https://www.youtube.com/watch?v=" + videoId + "&list=RDMM" + videoId);
    assertTrue(extractor instanceof YoutubeMixPlaylistExtractor);
    final String mixVideoId = "qHtzO49SDmk";
    extractor = YouTube.getPlaylistExtractor(
        "https://www.youtube.com/watch?v=" + mixVideoId + "&list=RD" + videoId);
    assertTrue(extractor instanceof YoutubeMixPlaylistExtractor);}
\end{minted}
\vspace{-5pt}

To study this phenomenon systematically, we analyzed tests that contain multiple call-assertion sequences. From each sequence, we extracted all non-assertion calls and compared them across sequences using two strategies. The first is \textit{order-independent matching}, which ignores call order and compares method signatures and receiver types. The second is \textit{order-dependent matching}, which is more restrictive and requires that both the set and order of calls match. We then categorized the matches by their degree of similarity. 
Our analysis shows that such repetition is common: in tests with more than one call-assertion sequence, 48.7\% share the same set of method calls and 48.0\% have identical sequences, including the order of calls. This suggests that developers frequently exercise similar execution paths with different inputs, reinforcing the role of call-assertion blocks as a recurring structural pattern in developer-written tests.

This pattern reveals an important insight: developers often consolidate repeated testing blocks—invoking the same methods with different inputs—into a single test method, even though these blocks could be expressed as separate tests. In contrast, most ATG tools typically generate such blocks as distinct test cases. While this consolidation increases the complexity of individual test methods, it provides practical benefits: related assertions remain co-located, simplifying test maintenance and allowing developers to evolve tests consistently as the code changes or new edge cases arise.

\begin{findingbox}{7}
{
Developer-written tests often contain repeated blocks that invoke the same methods with different inputs. In the \hamster dataset, when a test includes multiple call-assertion sequences, 48\% of them have an identical counterpart within the same method.
}
\end{findingbox}
\vspace{3pt}

\noindent
\textbf{Reusable helper methods.}
Helpers, which we previously defined as methods within the same test class and invoked in the test method, enhance modularity by encapsulating shared logic, such as repeated calls or assertions, thereby reducing duplication. We observed several test classes that define at least one helper method; within these, helpers are often invoked multiple times to standardize common setups or verifications.

For example, in the Adempiere project~\cite{adempiere}, the helper method \smalltt{setGardenWorldClientExists} creates a mocked \smalltt{MClient} object and delegates to another helper for mocking a \smalltt{Query}. These helpers are reused across tests to configure consistent test states. 

\vspace{-2pt}
\begin{minted}[frame=lines,framesep=1mm,baselinestretch=1, fontsize=\scriptsize, breaklines, breakanywhere, linenos,numbersep=4pt]{java}
final void ifGardenWorldIsUpToDate_ReturnsMsg() {
    setGardenWorldClientExists();
    doNothing().when(uut).clearSessionLog();...}
private void setGardenWorldClientExists() {
    MClient clientMock = mock(MClient.class);
    setQueryToReturnClient(clientMock);}
private void setQueryToReturnClient(MClient client) {
        Query queryMock = mock(Query.class);
        doReturn(queryMock).when(queryMock).setParameters(11); ...}
\end{minted}
\vspace{-2pt}

\noindent
\textit{Patterns and characteristics of test assertions.}
Assertions verify test outcomes, and their patterns reveal sophistication in developer practices. Across 1,697,196 tests with 6,922,089 assertions, we classified them into different types: 
equality assertions (23.9\%), assertion wrappers (21.0\%), truthiness assertions (16.7\%), string assertions (14.8\%), and collection assertions (8.2\%) are the most prevalent ones. 
Assertion wrappers---which are assertion entry-point methods (e.g., \smalltt{assertThat} in Hamcrest~\cite{hamcrest})---delegate to more specialized assertions and enable expressive composition.

\noindent
\textbf{Assertion chaining.}
Assertion chaining refers to the practice of linking multiple verification conditions together, encoding intricate logic and assertion semantics. Chained assertions begin with an assertion wrapper (e.g., \smalltt{assertThat}) applied to a target object, followed by a sequence of chained method calls.
For instance, in the following assertion from the Spring Cloud Gateway project~\cite{springcloud}, an assertion chain is used to first assert that the map is not null before asserting that it contains two key entries: \smalltt{assertThat(map).isNotNull().containsEntry().containsEntry()}.\\ Overall, 61,466 tests (3.6\%) contain assertion chains with two or more assertions on the wrapper, with an average chain length of 3.4 and a maximum of 358.

\noindent
\textbf{Grouped assertions.}
Grouped assertions (e.g., JUnit's \smalltt{assertAll}) execute all sub-checks regardless of any failures, aiding in the debugging process. We found 7,093 instances (<1\% of tests) of grouped assertions, such as this example from the AWS SDK project~\cite{aws-sdk-java-v2}:
\vspace{-4pt}
\begin{minted}[frame=lines,framesep=1mm,baselinestretch=1, fontsize=\scriptsize, breaklines, breakanywhere, linenos,numbersep=4pt,]{java}
assertAll(
() -> assertThatThrownBy(() -> client.noSigv4aPropertiesInEndpointRules(r -> r.stringMember(""))).hasMessageContaining("stop"),
() -> assertThat(signer.request.property(AwsV4aHttpSigner.REGION_SET)).isNull(),
() -> assertThat(signer.request.property(AwsV4aHttpSigner.SERVICE_SIGNING_NAME)).isEqualTo("fromruleset"));
\end{minted}
\vspace{-4pt}
Although less common and harder to implement, grouped assertions aid developers by streamlining troubleshooting.  

\noindent
\textbf{Assertions within helper methods.}
Assertions are often added to helper methods so multiple tests can reuse the same code. Our examination found that 175,248 (10.3\%) of the test cases employ helper methods for assertions.

\noindent
\textbf{Composite Arrange-Act-Assert.}
The Arrange-Act-Assert (AAA) pattern promotes separation among the setup, execution, and verification phases of tests to enhance readability~\cite{tdd_beck_2002}. The presence of multiple call-assertion blocks within a test method signals cases where the act phase is intertwined with assertions. We found that 758,034 tests (44.6\%) feature two or more such call-assertion blocks, suggesting the presence of a recurring variation we term the composite AAA framework---a blended structure that deviates from the rigid AAA model yet manifests frequently in real-world tests.

\noindent
\textbf{Under-testing and over-testing.}
We measured assertion density as percentages of callable entities and NCLOC for test methods. Our analysis reveals that, on average, 19.9\% (median 15.8\%) of entities are assertions, and 23.7\% (median 16.1\%) of NCLOC are assertions. The extrema cases indicate potential issues: the 25th percentile has 3.5\% entity density and 3.6\% line density, suggesting under-testing; the 90th percentile has 50.0\% entity density and 52.9\% line density, risking over-testing and brittleness. Notably, 372,939 tests (22.0\%) lack assertions, instead relying on print statements for manual inspection or thrown errors in the system under test. This observation is consistent with prior empirical findings~\cite{zamprogno2022dynamic, maayan_quality_2018}.

\begin{findingbox}{8}
{Developer-written tests employ a wide range of assertion types, often featuring complex patterns such as chained or grouped assertions, and achieving modularity through helper methods. At the same time, a notable share of tests suffer from extremes: some exhibit severe under-testing, containing few or no assertions, while others show over-testing, with excessively high assertion density.}
\end{findingbox}

\vskip 2pt
\noindent\textit{Comparison with ATG Techniques.} Figure~\ref{fig:testmethod} illustrates how ATG-generated tests differ from developer-written ones along different dimensions. Across most metrics, developer-written tests are significantly more complex, especially when exercising multiple focal classes. Developers also structure tests with repeated call-assertion blocks, employ helper methods to build call sequences and organize checks, and frequently group assertions to capture related properties and provide thorough failure reporting. In contrast, ATG-generated tests are far simpler: they almost never use helpers, avoid repeated call-assertion patterns, and rely instead on isolated, one-off assertions, resulting in lower overall assertion richness. The one notable exception is \aster with its mocking capabilities, as it frequently introduces multiple mock objects in its generated tests.

\vskip 2pt
\noindent\textit{Implications.}
These observed differences highlight several promising research directions for advancing ATG approaches. A central challenge is enabling ATG methods to capture the structural and semantic richness found in developer-written tests. This includes supporting complex assertion chains that validate multi-step behaviors, grouping related assertions to provide more comprehensive failure reporting, and introducing helper methods that promote modularity and reuse across test suites.
Beyond structural improvements, LLMs offer the potential to infer context-aware assertions grounded in code semantics, execution flows, and the overall intent of the test. This would enable ATGs to not only generate richer, more realistic tests, but also enhance developer-written tests by augmenting assertions when needed.

\begin{implicationbox}{3}
{
Future work should explore leveraging LLMs to incorporate assertion chains, modular test structures, and context-aware assertions into tool-generated tests. Another promising direction is the development of techniques that complement developer-written suites by detecting and addressing weaknesses such as missing or insufficient assertions.

}
\end{implicationbox}

\subsection{RQ5: Code Coverage}
\label{subsec:sota}
\begin{figure}[t]
    \centering
    \includegraphics[width=0.55\linewidth]{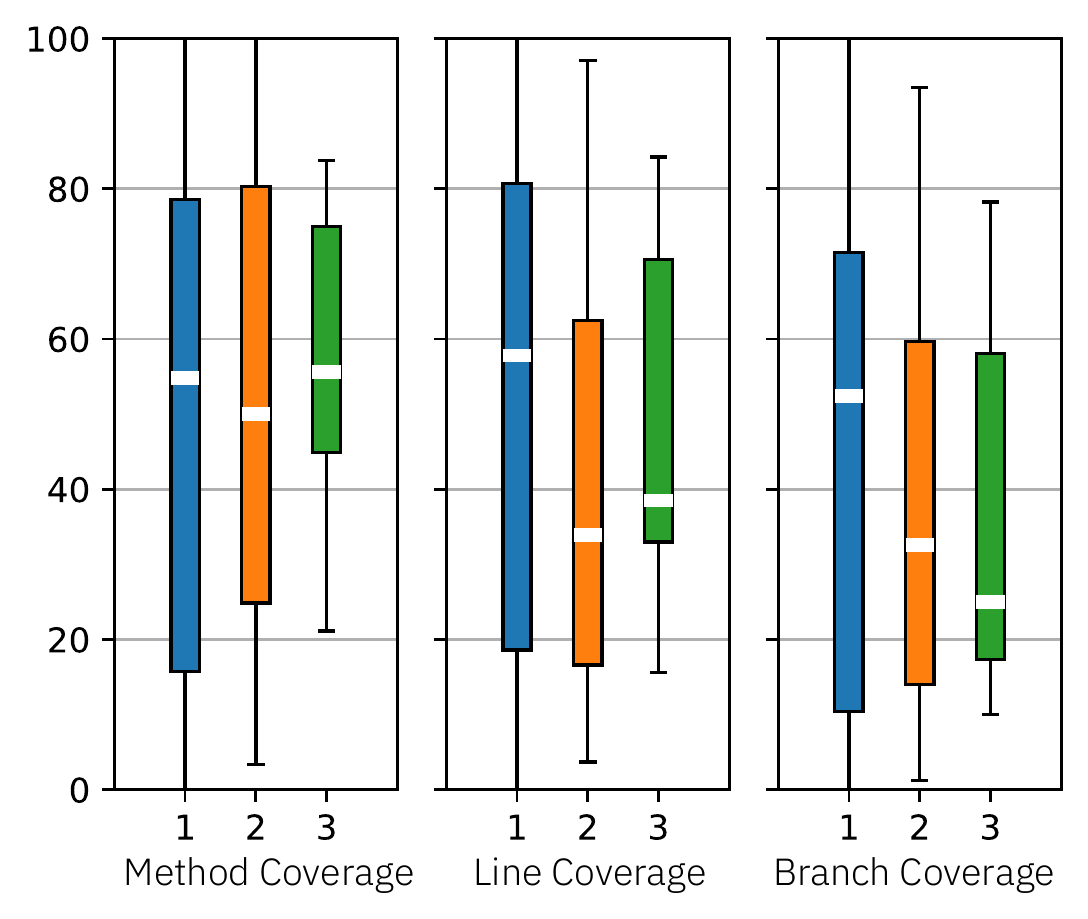}
    \vspace{-2pt}
    \caption{ Coverage achieved by developer-written and ATG-generated tests (1: developer, 2: EvoSuite, and 3: \aster).}
    \vspace{-2pt}
    \label{fig:sota}
\end{figure}

In addition to analyzing test characteristics, we also evaluated the code coverage achieved by developer-written and ATG-generated tests. Due to compilation and compatibility constraints, we considered a subset of the \hamster dataset: out of 1,908 repositories, 133 projects were successfully compiled, executed, and instrumented using the JaCoCo plugin~\cite{jacoco}. Among these, we computed coverage for EvoSuite-generated tests on 40 projects; for \aster, we used the tests reported in the original paper, spanning 7 projects and 6 LLMs. Figure~\ref{fig:sota} summarizes the results, showing that developer-written tests consistently achieve higher coverage than those generated by EvoSuite and \aster. This gap can be attributed to the broader scope of developer-written tests, which often extend beyond unit testing and capture more complex testing scenarios.

\vspace{-5pt}
\section{Threats to Validity}
\label{sec:threat}


\vskip 1pt
\noindent\textit{Construct threat.} 
A possible threat arises from our choice of metrics. To mitigate this, we relied on well-established measures widely used in prior work, such as NCLOC, cyclomatic complexity, number of test methods, and counts of method-call and assertion sequences. We also extended beyond prior work by introducing new analyses—focal classes, focal methods, and call–assertion patterns—that provide a richer view of test characteristics. Another potential threat is that some open-source Java projects in our dataset may already contain ATG-generated tests. However, the stark contrasts we observed between developer-written and tool-generated tests suggest that such cases are rare and unlikely to affect our findings.


\vskip 2pt
\noindent\textit{Internal threat.} The primary threat to internal validity concerns the accuracy of our static analysis. To mitigate this risk, we relied on well-established analysis engines, JavaParser and Tree-sitter~\cite{cldk}.


\vskip 2pt
\noindent\textit{External threat.} 
One potential threat to external validity is whether the test characteristics we observed generalize beyond the Java ecosystem. Although our study focuses on Java, we believe that many of the patterns we identified—such as test complexity, reliance on fixtures, and structured assertion blocks—are language-agnostic. Moreover, as Java benefits from some of the most mature ATG tools, the gap between developer-written and ATG-generated tests is likely even wider in less-supported languages, making our conclusions more compelling in those contexts. Importantly, both our analysis approach and the \hamster code model were designed to be language-independent and can be adapted to other programming languages with minimal effort.  
To mitigate potential threats to dataset quality, we curated a large and diverse collection of 1,908 high-quality Java projects comprising 1.7M test cases, selected using multiple criteria to reflect real-world development practices.


\vspace{-4pt}
\section{Related Work}
\label{sec:related}


Large-scale empirical studies have mined open-source repositories to characterize developer-written tests using static analysis. Kochhar et al.~\cite{kochhar_empirical_2013} analyzed 20,817 GitHub projects to collect metrics such as test counts, and subsequent work~\cite{kochhar_empirical_2014} explored relationships between test complexity and coverage. Other studies examine testing patterns like research on test smells from early detection and refactoring efforts~\cite{van2001refactoring, van2007detection, meszaros2007xunit} to tools like tsDetect~\cite{peruma_tsdetect_2020} and PyNose~\cite{wang_pynose_2021}. Survey-based studies~\cite{kurmaku_human-based_2022, fraser:2015:developerstudy, ceccato_automatically_2015, serra_effectiveness_2019} have provided further insights into developer challenges and practices.

More targeted analyses examine specific aspects of test cases. Assertion-based studies use assertion density as a proxy for test quality~\cite{kudrjavets2006assessing, athanasiou2014test}, while focal method identification links assertions to the methods under test~\cite{ghafari2015automatically}. These approaches help uncover test intent but are largely limited to unit tests.

Prior work on ATG tools often compares them with developer-written tests using metrics such as coverage, mutation scores, and bug detection~\cite{kurmaku_human-based_2022, fraser:2015:developerstudy, ceccato_automatically_2015, serra_effectiveness_2019, pan2025aster}. While useful, these studies leave some aspects of structural similarity less explored.

\hamster addresses these limitations with a comprehensive code model capturing fine-grained attributes to systematically characterize developer-written tests at scale, revealing testing variety beyond unit tests and exposing qualitative gaps in ATG-generated tests that traditional metrics overlook.

\vspace{-7pt}
\section{Summary and Future Work}
\label{sec:conclusion}

We presented a large-scale empirical study aimed at characterizing developer-written tests and identifying limitations of current ATG tools. Our analysis of 1.7M test cases reveals that developer-written tests exhibit characteristics that are often absent in tool-generated tests, such as exercising multiple methods and classes, using structured external inputs, containing sophisticated fixtures involving mocks, and employing complex assertions. ATG tools are typically limited to targeting individual methods and using simplistic input patterns. Our findings motivate further research along multiple directions: expanding Hamster to other programming languages; investigating how ATG tools can capture higher-level test intent through program analysis, specification mining, and machine learning; developing LLM-based techniques to mimic developer testing strategies (mocking, complex assertions, structured inputs); extending ATG capabilities to integration and system tests spanning multiple classes; and creating evaluation methodologies that go beyond coverage metrics and can assess the realism, maintainability, and alignment with developers' workflow of tests. In general, we believe this work provides an empirical foundation for developing ATG techniques able to generate tests that developers can integrate, maintain, and evolve alongside their codebases.



\bibliographystyle{ACM-Reference-Format}
\bibliography{references}

\end{document}